\documentclass[10pt,A4paper,conference]{IEEEtran}
\IEEEoverridecommandlockouts
\usepackage{cite}
\usepackage{amsmath,amssymb,amsfonts}
\usepackage{graphicx}
\usepackage{textcomp}
\usepackage{xcolor}
\usepackage{tabularx}
\usepackage{color, colortbl}
\usepackage[misc]{ifsym}

\definecolor{lime}{rgb}{0.88,2,10}
\usepackage[left=.63in,
            right=.63in,
            top=.66in,
            bottom=0.94in]{geometry}

\renewcommand{\baselinestretch}{.91}
\usepackage{subcaption}
\usepackage{wrapfig}
\usepackage[linesnumbered,ruled,vlined]{algorithm2e}
\usepackage{xpatch}
\usepackage{url}
\usepackage{multirow}
\usepackage{multicol}
\usepackage{wrapfig}

\usepackage[export]{adjustbox}
\usepackage{nomencl}
\usepackage{siunitx}
\usepackage{blindtext}
\usepackage[T1]{fontenc}
\usepackage[spaces,hyphens]{xurl}
\usepackage{float}

\newcommand*{\Resize}[2]{\resizebox{#1}{!}{$#2$}}%

\SetAlFnt{\small}

\SetKwComment{Comment}{$\triangleright$\ }{}

\usepackage[utf8]{inputenc}
\usepackage{enumitem}

\SetKwInput{kwInit}{Initialize}

\def\BibTeX{{\rm B\kern-.05em{\sc i\kern-.025em b}\kern-.08em
    T\kern-.1667em\lower.7ex\hbox{E}\kern-.125emX}}

\usepackage[numbers,sort&compress]{natbib}
\usepackage[font={small}]{caption} 

\newcommand{\fref}[1]{Fig.~\ref{#1}}

\newcommand{\sref}[1]{Section~\ref{#1}}
\usepackage[misc]{ifsym}

\usepackage[outline]{contour}
\contourlength{0.2pt}
\contournumber{15}
\newcommand{\mycontour}[2][black!90]{\textcolor{white}{\contour{#1}{#2}}}
\newcommand{\mycontourr}[2][black!70]{\textcolor{black!70}{\contour{#1}{#2}}}
\newcommand{\mycontourrr}[2][yellow]{\textcolor{yellow}{\contour{#1}{#2}}}
\newcommand\HUGE{\fontsize{18.95}{25}\selectfont}

\usepackage{fancyhdr}
\fancypagestyle{mystyle}{
    \chead{\large 2023 IEEE Global Communications Conference (GLOBECOM 2023)}}

\begin{document}
\title{\HUGE {Secure and Efficient Federated Learning in LEO Constellations using Decentralized Key Generation and On-Orbit Model Aggregation} \vspace{-0.1cm}}

\author{\IEEEauthorblockN{
 Mohamed Elmahallawy\IEEEauthorrefmark{1}, Tie Luo\IEEEauthorrefmark{1}\IEEEauthorrefmark{2}, Mohamed I. Ibrahem\IEEEauthorrefmark{3}}
      \IEEEauthorblockA{%
  \IEEEauthorrefmark{1}Computer Science Department, Missouri University of Science and Technology, Rolla, MO 65401, USA}
  \IEEEauthorblockA{%
    \IEEEauthorrefmark{3}School of Computer and Cyber Sciences, Augusta University, Augusta, GA 30912, USA}
\thanks{\IEEEauthorrefmark{2}Corresponding author. This work is supported by the National Science Foundation (NSF) under Grant No. 2008878.}  
  Emails:  \{meqxk, tluo\}@mst.edu, mibrahem@augusta.edu

}

\maketitle
\thispagestyle{mystyle}
\begin{abstract}
Satellite technologies have advanced drastically in recent years, leading to a heated interest in launching small satellites into low Earth orbit (LEOs) to collect massive data such as satellite imagery. Downloading these data to a ground station (GS) to perform centralized learning to build an AI model is not practical due to the limited and expensive bandwidth. Federated learning (FL) offers a potential solution but will incur a very large convergence delay due to the highly sporadic and irregular connectivity between LEO satellites and GS. In addition, there are significant security and privacy risks where eavesdroppers or curious servers/satellites may infer raw data from satellites' model parameters transmitted over insecure communication channels. To address these issues, this paper proposes {\em FedSecure}, a secure FL approach designed for LEO constellations, which consists of two novel components: (1) \textit{decentralized key generation} that protects satellite data privacy using a functional encryption scheme, and (2) \textit{on-orbit model forwarding and aggregation} that generates a partial global model per orbit to minimize the idle waiting time for invisible satellites to enter the visible zone of the GS. Our analysis and results show that FedSecure preserves the privacy of each satellite's data against eavesdroppers, a curious server, or curious satellites. It is lightweight with significantly lower communication and computation overheads than other privacy-preserving FL aggregation approaches. It also reduces convergence delay drastically from days to only a few hours, yet achieving high accuracy of up to 85.35\%  using realistic satellite images.
\end{abstract}

\begin{IEEEkeywords} Low Earth orbit (LEO), satellite communication (SatCom), federated learning (FL), privacy preservation. \end{IEEEkeywords}

\section{Introduction}\label{Intro}

{\bf Background.} The advancement of satellite technology has enabled the launching of many small satellites into low Earth orbits (LEOs). Equipped with multiple sensors and cameras, these satellites gather extensive data about the Earth and space, allowing large AI models to be trained to support various applications such as monitoring remote areas like deserts, forests, and maritime regions, as well as homeland security like border surveillance and military reconnaissance. However, the traditional approach of centralized learning, which requires downloading the satellite data (e.g., imagery) to a ground station (GS), is impractical due to the limited bandwidth, highly intermittent connectivity between satellites and GS, and data privacy.

Federated learning (FL) \cite{mcmahan2017communication} offers a promising solution as a distributed learning paradigm, which both saves bandwidth and preserves privacy. In FL, each client (satellite in our context) trains a local machine learning (ML) model onboard and sends only the model parameters (instead of raw data) to an aggregation server ${\mathcal {AS}}$ (GS in our context). The ${\mathcal {AS}}$ then aggregates all the received local models into a global model and sends it back to all the satellites for re-training. This procedure repeats until the global model eventually converges.

\textbf{FL-LEO Challenges.} However, applying FL to satellite communications (SatCom), or more specifically LEO constellations, faces significant challenges. First, there is a large delay in every communication round between satellites and ${\mathcal {AS}}$ and it leads to a very slow FL convergence process which often takes several days \cite{chen2022satellite}. This delay is caused by the highly irregular and sporadic connectivity between satellites and the ${\mathcal {AS}}$, which is attributed to the distinct Earth rotation and satellite orbiting trajectories. Second, transmitting model parameters in FL may appear ``safe'' but is in fact still vulnerable under certain attacks such as model inversion and membership inference \cite{nasr2019comprehensive}. 

\textbf{Contributions.} 
To address the above challenges, we propose FedSecure, a secure FL-LEO framework to ensure both fast convergence and protection of privacy leakage, while maintaining high accuracy of the global model. Specifically,
\begin{itemize}[leftmargin=*]

\item FedSecure consists of a functional encryption-based aggregation scheme that prevents the private satellite data from being inferred and satellite models from being deciphered, {\em without requiring any trusted key distribution center (KDC) to generate public/private keys or any secure channel to distribute the keys among nodes}.

\item We also propose an on-orbit model forwarding and aggregation scheme that generates a {\em partial global model} per orbit via intra-orbit collaboration, which significantly reduces the waiting time for invisible satellites to enter the visible zone of ${\mathcal {AS}}$ for model transmissions. 

\item Our extensive experiments on a real satellite imagery dataset for semantic segmentation tasks demonstrate that FedSecure achieves convergence in only 3 hours while maintaining competitive accuracy across various performance metrics including IoU and the Dice Coefficient. It is also lightweight with low computation overhead ($<9$ ms) and communication overhead (497 MB).
\end{itemize}
\section{Related Work}\label{related}
Despite the relative youth of the FL-LEO research field, notable studies have made initial strides in this area \cite{razmi, eloptimizing, lin2022federated, elfedhap, elAsyn, so2022fedspace, wang2022fl, elmahallawy2023one}. In the synchronous FL category \cite{razmi, eloptimizing, lin2022federated, elfedhap}, the ${\mathcal {AS}}$  waits to receive {\em all} the satellites' models in each training round. FedISL \cite{razmi} uses inter-satellite-link (ISL) to reduce the waiting time, but it achieves fast convergence only when the ${\mathcal {AS}}$ is a GS located at the north pole (NP) or a satellite in a medium Earth orbit (MEO) above the Equator, otherwise it needs several days for convergence. The work \cite{eloptimizing} removes these restrictions and, in order to reduce delay, designates a sink satellite per orbit to collect models from satellites in the same orbit. However, it requires each satellite to run a distributed scheduler which incurs extra delay.  In \cite{lin2022federated}, the authors proposed a method for dynamically aggregating satellite models based on connection density, involving multiple GSs collaboration. However, ensuring model consistency across GSs is challenging and adds overhead. Lastly, the authors of \cite{elfedhap} proposed FedHAP which uses multiple airships or balloons to act as ${\mathcal {AS}}$s to collect models from satellites, but it requires extra hardware (HAPs) to be deployed.

In the asynchronous FL category \cite{elAsyn, so2022fedspace, wang2022fl}, the ${\mathcal {AS}}$ only collects models from a subset of satellites in each training round. One such approach is AsyncFLEO \cite{elAsyn} which groups satellites according to model staleness and selects only fresh models from each group while down-weighting outdated models. So et al. proposed FedSpace \cite{so2022fedspace}, which aims to balance the idle waiting in synchronous FL and the model staleness in asynchronous FL by scheduling the aggregation process based on satellite connectivity. However, it requires satellites to upload a portion of raw data to the GS, which contradicts the FL principles on communication efficiency and data privacy.
In \cite{wang2022fl}, a graph-based routing and resource reservation algorithm is introduced to optimize the delay in FL model parameter transfer. The algorithm improves a storage time-aggregated graph, providing a comprehensive representation of the satellite networks' transmission, storage, and computing resources.


To the best of our knowledge, our work is the first that addresses security threats in FL-LEO against both internal and external adversaries (cf. \sref{Threat}). Although some existing privacy-preserving and cryptographic techniques such as homomorphic encryption \cite{ma2022privacy} 
could be applied, these approaches have limitations such as large ciphertexts and high communication overhead; 
also importantly, they require a trusted KDC to generate and distribute public and private keys which further requires a secure communication channel as well between all clients and the ${\mathcal {AS}}$. Other classical cryptographic protocols such as differential privacy (DP) and secure multi-party computation (SMC) can also be applied, but they suffer from high encryption and communication overheads and can degrade the accuracy of the global model (e.g., DP adds noise to local models). 
\begin{figure}[t] 
     \centering    \includegraphics[width=0.8\linewidth,height=6cm]{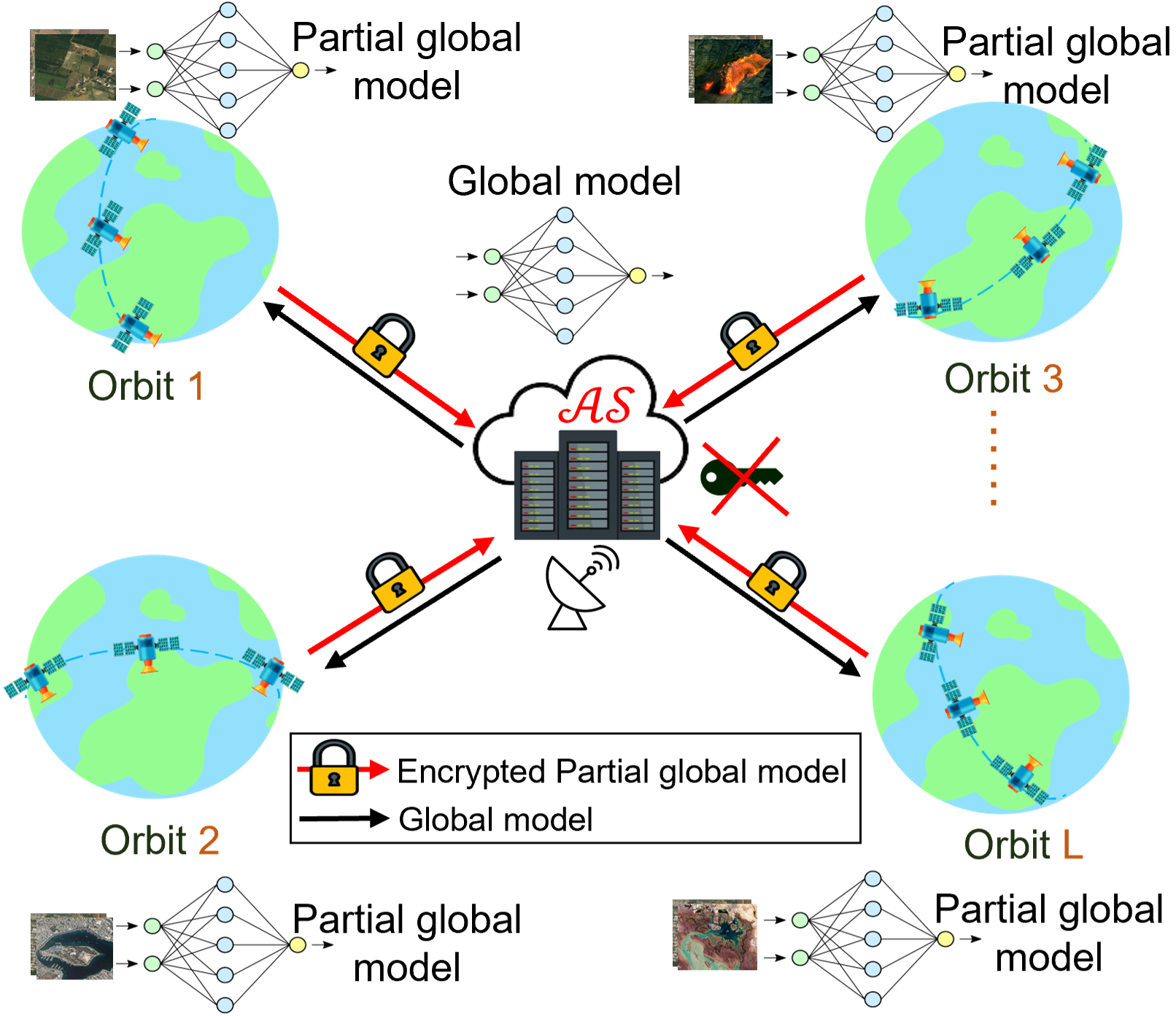}
    \caption{System model: an LEO constellation with multiple (4) orbits.}
    \label{system_model}
    \vspace{-0.5 cm}
\end{figure}

 \vspace{-0.15 cm}
\section{System and Threat Model} \label{sec:model}
Our objective is to develop a decentralized FL-LEO approach that ensures both data and model privacy of LEO satellites, without requiring a KDC (and the associated secure channel). 

\subsection{System Model}
Our system model (\fref{system_model}) comprises:

\begin{enumerate}[label=\arabic*)]
    
\item LEO satellites: An LEO constellation $\mathcal K$ consists of multiple satellites, indexed by $i$, orbiting the Earth in ${L}$ orbits. Each orbit $l$ has a set of equally-spaced satellites. While orbiting, each satellite $i$ captures high-resolution images for training an ML model for various classification tasks (e.g., detecting forest fires or hurricanes, monitoring country borders, etc.). During each round of FL training, all LEO satellites receive an (initial or updated) global model from the ${\mathcal {AS}}$ during their respective visible windows, and then (re)train the model using their own local data. After training, they encrypt the model parameters and send them back to the ${\mathcal {AS}}$, which aggregates them into a global model and then sends it back to all satellites again.

\item Aggregation Server $\mathcal {AS}$: The ${\mathcal {AS}}$ initiates the FL process by sending an initial (typically randomly initialized) global model to all LEO satellites successively during their respective visible windows. Then, after receiving the retrained local models from the visible satellites successively, the ${\mathcal {AS}}$ aggregates them into an updated global model, and broadcasts it back to visible satellites again for another round of training. This process continues until a termination criterion is met, such as reaching a target accuracy or loss, maximum number of communication rounds, or negligible change of model parameters.

\end{enumerate}

\subsection{Threat Model}\label{Threat}
Our threat model encompasses both external and internal adversaries, as outlined below:
\begin{enumerate}
    \item {\bf External adversaries.} An adversary may eavesdrop on the communication links between LEO satellites and the ${\mathcal {AS}}$ to steal or monitor the model parameters, thereby inferring sensitive information about satellite data from their parameters.

    \item {\bf Internal adversaries.} The ${\mathcal {AS}}$ and LEO satellites are {\em honest-but-curious} participants in FL training, meaning that they follow the FL protocol honestly but may be curious to learn/infer sensitive information about the raw data of some or all the satellites. In addition, we also consider possible collusion between the ${\mathcal {AS}}$ and some satellites (e.g., launched by an operator, to steal information from satellites owned by another operator).
\end{enumerate}

In general, the transmission of local model parameters may leave LEO satellites vulnerable to attacks such as membership inference, model inversion, etc., which are all subsumed by the above threat model.

\subsection{Design Goals}
\begin{enumerate}
    \item {\bf Privacy preservation:} The solution should ensure the confidentiality of the ML model parameters and prevent any information leakage to attackers/eavesdroppers, the ${\mathcal {AS}}$, or other LEO satellites.
    
    \item {\bf Efficiency:} The available communication bandwidth should be used efficiently. This may involve minimizing the model size and exchange frequency between the ${\mathcal {AS}}$ and satellites, as well as reducing communication overhead.

    \item {\bf Accuracy:} The final global model resulting from FL should still achieve competitive performance.
\end{enumerate}

\section{Federated Learning in LEO constellations}

\subsection{FL-LEO's Computation Model}

The overarching objective of the ${\mathcal{AS}}$ and all LEO  satellites $\mathcal{K}$ is to collaboratively train a global ML model, which involves the following steps: (i) the ${\mathcal{AS}}$ initializes an ML model and sends it to each individual satellite when that satellite enters the ${\mathcal{AS}}$' visible zone; (ii) each satellite then trains the model using a local optimization method, typically mini-batch gradient descent, $\boldsymbol {w}_{i}^{\beta,j+1} = \boldsymbol {w}_{i}^{\beta,j}- \zeta \nabla F_{k}(\boldsymbol {w}_{i}^{\beta,j};X_{i}^{j})$, where $\boldsymbol {w}_{i}^{\beta,j}$ is the local model of satellite $i$ at the $j$-th local iteration in a global communication round $\beta$, $\zeta$ is the learning rate, $X_{i}^{j}\subset D_{i}$ is the $j$-th mini-batch and $D_{i}$ is satellite $i$'s dataset; after training, sends the updated model $\boldsymbol {w}_{i}^{\beta,J}$ back to the ${\mathcal{AS}}$; (iii) once the ${\mathcal{AS}}$ receives the trained models from all satellites, it aggregates them into an updated global model $\boldsymbol {w}^{\beta+1} = \sum_{i\in \mathcal K} \boldsymbol {w}_{i}^{\beta,J}$ and sends it back to all satellites during their respective visible windows as in (i). The above process continues until the global model converges.


\subsection{FL-LEO's Communication Model}\label{Com_link}
Assuming line-of-sight (LoS) communication, a satellite $i$ and an $\mathcal{AS}$ $s$ can only communicate with each other if $\angle (r_{s}(t), (r_{i}(t) - r_{s}(t))) \leq \frac{\pi}{2}-\alpha_{min}$, where $r_{i}(t)$ and $r_{s}(t)$ are their respective trajectories and $\alpha_{min}$ is the minimum elevation angle. Additionally, assuming the channel is affected by additive white Gaussian noise (AWGN), the signal-to-noise ratio (SNR) between them is $\frac{P G_{i}G_{s}}{K_{B} T B \mathcal{L}_{i,s}}$ where $P$ is the transmitter power, $G_{i}, G_{s}$ are the antenna gain of $i$ and $s$, respectively, $K_{B}$ is the Boltzmann constant, $T$ is the noise temperature, \textit{B} is the channel bandwidth, and $\mathcal{L}_{i,s}$ is the free-space pass loss which can be calculated as $\mathcal{L}_{i,s} = \big(\frac{4\pi \|i,s\|_{2} f}{c}\big)^{2}$. Here,  $\|i,s\|_{2} $ is the Euclidean distance between ${i}$ and $s$ that satisfies $\|i,s\|_{2} \leq \mathnormal{\ell_{i,s}}$, where $\mathnormal{\ell_{i,s}}$ is the minimum distance between ${i}$ and $s$ that enables them to communicate with each other, and $f$ is the carrier frequency. 


\subsection{Inner-Product Functional Encryption}

Unlike conventional encryption methods, functional encryption (FE) is a cryptosystem that allows the holder of a decryption key to decrypt encrypted data but only obtain a {\em function} of the data without revealing the input data itself \cite{abdalla2018multi}. Our work focuses on {\em inner product FE} (IPFE) which is a specific type of FE that performs a function of inner product operations over encrypted data \cite{abdalla2018multi,kim2018function}. With IPFE, given two ciphertext vectors $\mathbf{x}'$ and $\mathbf{y}'$ which are encrypted from plaintext vectors $\mathbf{x}$ and $\mathbf{y}$, one can obtain the inner product $\langle \mathbf{x}, \mathbf{y} \rangle$ without decrypting $\mathbf{x}'$ or $\mathbf{y}'$ or knowing any elements of $\mathbf{x}$ and $\mathbf{y}$. Compared to homomorphic encryption (HE) which requires decrypting the ciphertext to obtain the plaintext result, IPFE can directly obtain the final plaintext result.

\section{FedSecure Design} \label{sec:fedsec}
\subsection{Overview}\label{sec:overview}
FedSecure is a synchronous FL approach designed for LEO satellites to protect their data privacy while achieving high accuracy and speeding up convergence. It works as follows: (i) Each participating satellite $i \in \mathcal K$ generates a private key and a public key using the anonymous veto network (AV-net) protocol \cite{hao2009power}; (ii) The $\mathcal{AS}$ broadcasts the initial global model to all visible LEO satellites; (iii) Each visible satellite forwards the received global model to its neighbors on the same orbit so that invisible satellites can have the global model without waiting for their respective visible windows (see Fig.~\ref{On_Orbit_model}a); 
(iv) Once receiving the model, each satellite retrains it using its local data, and after training, encrypts the trained model's parameters using the AV-net protocol (\sref{sec:secagg}); (v) All the satellites on the same orbit collaborate to generate a partial global model by averaging their models and then encrypt and send this partial model to a currently visible satellite on the same orbit (\sref{onorbit_model}); 
(vi) This visible satellite per each orbit sends the ciphertext of the partial model to the $\mathcal{AS}$; (vii) The $\mathcal{AS}$ averages the encrypted parameters to obtain an updated global model {\em without being able to read any satellite's local model parameters or infer any satellite's training data} (\sref{sec:secagg}). A visualization of our on-orbit model forwarding and aggregation is shown in Fig.~\ref{On_Orbit_model}.


\subsection{On-Orbit Model Forwarding and Aggregation} \label{onorbit_model}
\begin{figure} [h]\vspace{-0.1cm}
     \centering
     \includegraphics[width=0.8\linewidth]{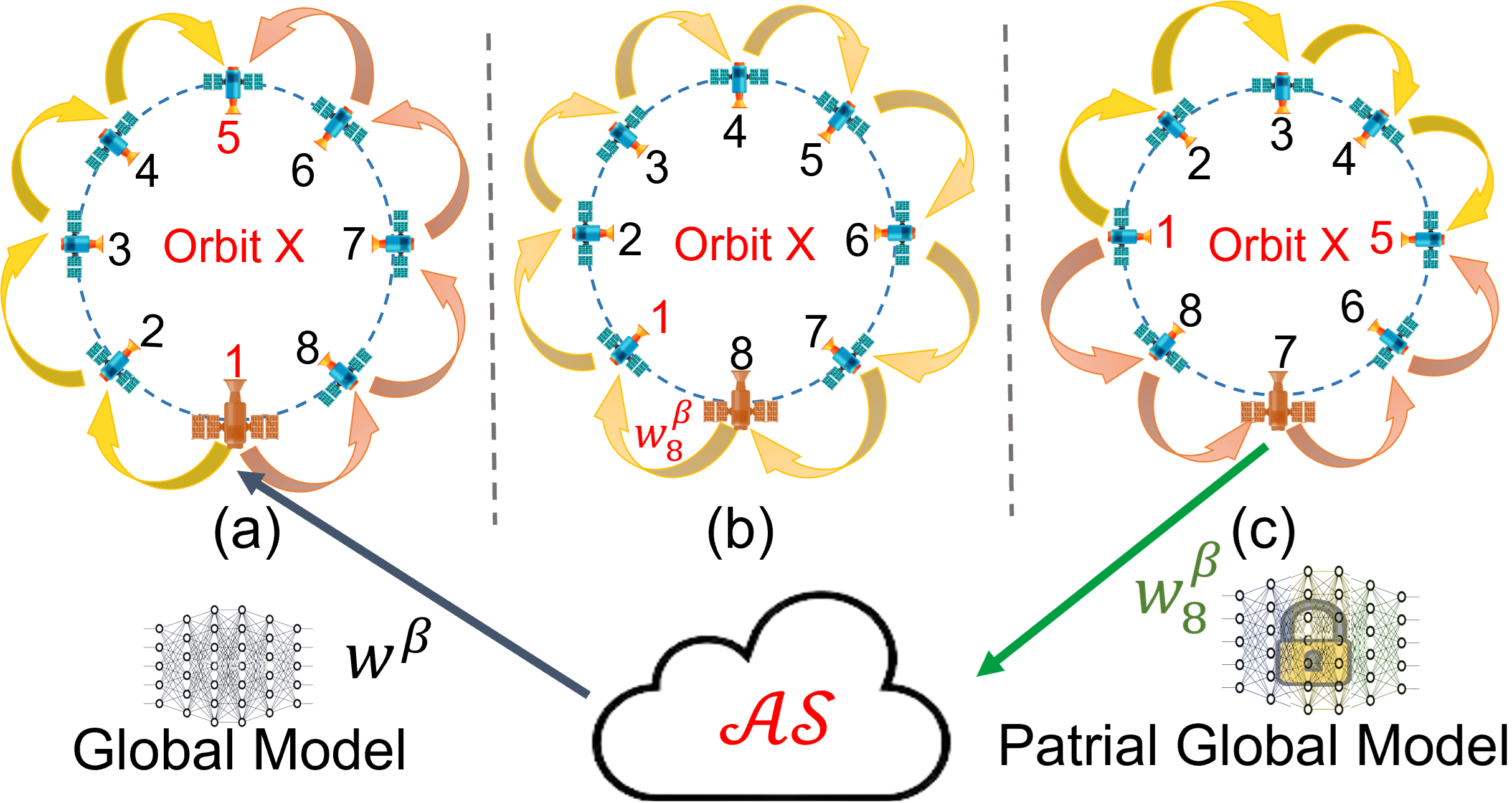}
    \caption{On-orbit model forwarding and aggregation. (a) The visible satellite (\#1 in the example) broadcasts the received $\boldsymbol {w}^\beta$ to all other satellites (most are invisible) on the same orbit bi-directionally; (b) it initiates model forwarding and aggregation by sending its trained local model $\boldsymbol {w}_1^\beta$ to its neighbor satellite uni-directionally (either clockwise or counterclockwise, predetermined); (c) after receiving the final updated partial model from \#8, it forwards the final partial model to all satellites on the same orbit bi-directionally, until reaching a visible satellite ($\#7$ in the example).}
    \label{On_Orbit_model}
    \vspace{-0.2 cm}
\end{figure}

We propose an on-orbit model forwarding and aggregation scheme to merge all satellites' local models on the same orbit into a {\em partial global model}. The purpose is to overcome the long idle waiting time for invisible satellites to (successively) enter the $\mathcal {AS}$'s visible zone for model exchange, as in traditional synchronous FL-LEO approaches \cite{chen2022satellite}. Our proposed scheme works as follows: (1) Each satellite $i$ on the same orbit trains the global model $\boldsymbol {w}^\beta$ received as per steps (ii) and (iii) of \sref{sec:overview}, and obtains $\boldsymbol {w}^\beta_{i}$ after training.
(2) Then, the first visible satellite, say \#1, forwards its $\boldsymbol {w}_1^\beta$ to its next neighbor (\#2, which may be invisible, see Fig.~\ref{On_Orbit_model}b), who will perform partial aggregation of its own $\boldsymbol {w}_2^\beta$ and the received $\boldsymbol {w}_1^\beta$ into a new $\boldsymbol {w}_2^\beta$, and passes it onto its next neighbor \#3; this process continues until the final partial model $\boldsymbol {w}_8^\beta$ reaches the originating satellite \#1. 
(3) Finally, satellite \#1 will forward this partial model back to all satellites on the same orbit (but this round without aggregation) so that any satellite who becomes visible the first will send that model to the $\mathcal {AS}$, for later global aggregation among all orbits. See \fref{On_Orbit_model}; more details are provided below.

Our on-orbit model forwarding and aggregation scheme is inspired by the method proposed in \cite{eloptimizing} but differs from it as follows. In \cite{eloptimizing}, each orbit needs to {\em schedule a sink satellite} to collect all local models and then perform partial aggregation, which involves both scheduling overhead and the waiting time for the sink satellite to become visible. But in our scheme here, every satellite participates in partial model aggregation without relying on a specific satellite, and hence anyone can send the final partial model to $\mathcal {AS}$. This makes our aggregation scheme more time-efficient (the additional round of bi-directional forwarding as in \fref{On_Orbit_model}c is very fast, taking a few seconds only).

Upon receiving the global model $\boldsymbol {w}^{\beta}$ generated by the $\mathcal{AS}$ for a global round $\beta$, a visible satellite $i$ in an orbit $X$ initiates a retraining process using its collected data (i.e., Earth imagery) to update its local model $\boldsymbol {w}_i^{\beta}$. After updating its local model, it forwards $\boldsymbol {w}^{\beta}$ to all satellites in its orbit, including invisible satellites. Subsequently, it transmits its updated local model $\boldsymbol {w}_{i}^{\beta}$ to its next-hop satellite $i'$ via Intra-plane ISL, with the propagation direction pre-designated either clockwise or counterclockwise. Then, the next-hop satellite $i'$ performs partial model aggregation by combining its updated local model $\boldsymbol {w}_{i'}^{\beta}$  with the received $\boldsymbol {w}_i^{\beta}$ to generate a new local model $\boldsymbol {w}_{i'}^{\beta}$ before transmitting it to $i''$ (the next-hop satellite in the designated propagation direction) as follows:
\begin{align}\label{eq:partialagg}
    \boldsymbol {w}_{i'}^{\beta}= (1-\alpha_{i'}) \boldsymbol {w}_{i}^{\beta} +  \alpha_{i'} \boldsymbol {w}_{i'}^{\beta}
\end{align}
where $\alpha_{i'}$ is a scaling factor, defined as the ratio between the data size of satellite $i'$ and the sum of all previous satellites (up to $i$)'s data size (each satellite will send its data size as metadata to its neighbor together with its model). Consequently, the visible satellite, say \#1, who initiated the above model forwarding process, will receive an updated on-orbit aggregated partial model $\boldsymbol {w}_{k}^{\beta}$ that has aggregated all the local models on the same orbit, where $k$ is the total number of satellites on that orbit (see \fref{On_Orbit_model}b). If the satellite \#1 is still visible to the $\mathcal{AS}$, it will send $\boldsymbol {w}_{k}^{\beta}$ to the $\mathcal{AS}$. Otherwise, it initiates another round of forwarding (but without aggregation) by sending $\boldsymbol {w}_{k}^{\beta}$ to its next-hop neighbors which will pass it onward further until either (i) $\boldsymbol {w}_{k}^{\beta}$ reaches a visible satellite, who will then send the model to the $\mathcal{AS}$ immediately, or (ii) every satellite on the same orbit has a copy of $\boldsymbol {w}_{k}^{\beta}$ and the first satellite who becomes visible will send the model to the $\mathcal{AS}$. 

To summarize, our on-orbit model forwarding and aggregation scheme eliminates the requirement of each satellite individually communicating with the $\mathcal {AS}$ as in conventional FL-LEO approaches, and thus cuts down the substantial idle waiting time for visible windows. Since this is entailed in {\em every} communication round of FL training, our scheme will lead to a significant acceleration of FL convergence.

\subsection{Decentralized Secure Aggregation Scheme}\label{sec:secagg}

Our novel secure model aggregation scheme allows building a global model without the need for a KDC while protecting the privacy of satellites. It encompasses three stages: system setup, training and reporting partial model parameters, and global parameter averaging. 

\begin{table}[ht]
	\centering
	\caption{System parameters.}
	\label{tab:multi_notation}
	\begin{tabular}{cl}
		\hline
		Parameter			& 	Description    \\
		\hline
	     
    \\[-0.7em]	
$\mathbb{G}$				&	Multiplicative cyclic group	with prime order $q$		            \tabularnewline \\[-0.7em]		
  $\mathrm{g}\in \mathbb{G}$				&	Randomly chosen generator		        \tabularnewline \\[-0.7em]
        $\mathbb{Z}_q$		&	Finite field of order $q$				    \tabularnewline \\[-0.7em]
		 $\mathcal{H}: \{0,1\}^* \rightarrow \mathbb{Z}_q$    & Full-domain hash function onto $\mathbb{Z}_q$   		      \tabularnewline \\[-0.7em]
        \hline
	\end{tabular}
	   \vspace{-1mm}
\end{table}
\subsubsection{System setup}
Pertaining to our discussion in Section~\ref{onorbit_model}, one visible satellite in each of the $L$ orbits (e.g., the first visible one) generates its public and secret keys 
using the AV-net protocol~\cite{hao2009power}. Given the system parameters in Table~\ref{tab:multi_notation}, the following steps are carried out:
\begin{enumerate}[leftmargin=*,label=(\alph*)]
\item The visible satellite $i_l$ chooses a private number $\mathrm{x}_{i_l}$ $\in \mathbb{Z}_q$ randomly, where $i_l$ is the visible satellite $i$ on orbit $l$.

\item Then, satellite $i_l$ broadcasts $\mathrm{g}^{\mathrm{x}_{i_l}}$ to all other visible satellites on different orbits (e.g., through the $\mathcal{AS}$ or the Internet), where $\mathrm{g}^{\mathrm{x}_{i_l}}$ is a public parameter. Note that this is only needed once in the initialization phase.

\item Each satellite $i_l$ computes $\mathrm{g}^{\mathrm{y}_{i_l}}$ as follows: \vspace{-0.06in}
        \begin{equation}
\mathrm{g}^{\mathrm{y}_{{i_l}}}=\prod_{\mathrm{z}=1}^{i_{l-1}} \mathrm{~g}^{\mathrm{x}_{\mathrm{z}}} / \prod_{\mathrm{z}={i_{l+1}}}^{i_L} \mathrm{g}^{\mathrm{x}_{\mathrm{z}}},
        \end{equation}

\item A secret key $s_{i_l} \in$ $\mathbb{Z}_{q}$ is chosen by each satellite $i_l$ to be used in the computation of the subset aggregation key $S_{ {i_l}}$ that will be sent to the $\mathcal{AS}$. The $S_{ {i_l}}$ is computed as follows:
    \vspace{-0.06in}
    \begin{equation}
    S_{{i_l}}=\mathrm{g}^{s_{i_l}}\times (\mathrm{g}^{\mathrm{y}_{i_l}})^{\mathrm{x}_{i_l}} = \mathrm{g}^{s_{i_l}+{\mathrm{x}_{{i_l}}\mathrm{y}_{{i_l}}}}\vspace{-0.05in}
    \end{equation}
    
\item After receiving $S_{ {i_l}}$ from all satellites, the $\mathcal{AS}$ computes an aggregation key $AK_\mathcal{AS}$, which will be used in the aggregation process as
\begin{align}
AK_\mathcal{AS} &= \prod_{i_{l=1}}^{i_L} S_{ {i_l}}=\prod_{i_{l=1}}^{i_L} \mathrm{g}^{s_{i_l}+{\mathrm{x}_{{i_l}}\mathrm{y}_{{i_l}}}}\nonumber \\
&= \mathrm{g}^{s_{i_1}+{\mathrm{x}_{{i_1}}\mathrm{y}_{{i_1}}}} \times \mathrm{g}^{s_{i_2}+{\mathrm{x}_{{i_2}}\mathrm{y}_{{i_2}}}} \times...\times
    \mathrm{g}^{s_{i_L}+{\mathrm{x}_{{i_L}}\mathrm{y}_{{i_L}}}}\nonumber \\
&=\mathrm{g}^{\sum_{{i_{l=1}}}^{i_L} s_{{i_l}}+\sum_{{i_{l=1}}}^{i_L} {\mathrm{x}_{i_l}\mathrm{y}_{i_l}}}
\end{align}
\end{enumerate}
Since $\sum_{{i_{l=1}}}^{ i_L} {\mathrm{x}_{i_l}\mathrm{y}_{i_l}}$=$~0$ as verified in~\cite{hao2009power}, $AK_\mathcal{AS} = \mathrm{g}^{\sum_{i_{l=1}}^{i_L} s_{i_l}}$.\\
\subsubsection{Secure reporting of partially trained models} \label{sub:Report Generation}
After a visible satellite obtains the partial global model $\boldsymbol{w}_{i_l}^\beta[\mu]$ in the $\beta$-th FL round as in \fref{On_Orbit_model}c, this satellite uses $s_{i_l}$ to encrypt this partial model's parameters before sending to the $\mathcal{AS}$:
\begin{equation}\label{Enc_eq}
        C_{i_l}^\beta[\boldsymbol{w}[\mu]]=\mathrm{g}^{{s_{i_l}{u_{\ell_\beta}}}+\boldsymbol{w}_{i_l}^\beta[\mu]} \in \mathbb{G}, \quad \mu=0,...,e-1
  \end{equation}
where $C_{i_l}^\beta$ is the ciphertext of model $\boldsymbol{w}_{i_l}^\beta[\mu]$ which contains $e$ parameters represented by the vector $(\boldsymbol{w}_{i_l}^\beta[0],\ldots ,\boldsymbol{w}_{i_l}^\beta[e-1])$,  $\ell_\beta$ is a round identifier, and ${u}_{\ell_\beta}=\mathcal{H}(\ell_\beta) \in \mathbb{Z}_q$.

\subsubsection{Global parameter-averaging}
This step is performed by the $\mathcal{AS}$. In each FL round $\beta$, the $\mathcal{AS}$ collects all the encrypted partial models from visible satellites on all the orbits, and then performs the following computation to calculate an aggregated model (that is still encrypted):
\begin{align}\Resize{8.2cm} {\frac{\prod_{{i_{l=1}}}^{i_L}(C_{i_l}^\beta[\mu])}{(AK_\mathcal{AS})^{u_{\ell_\beta}}} = \frac{\prod_{{i_{l=1}}}^{i_L}(\mathrm{g}^{{s_{i_l}{u_{\ell_\beta}}}+\boldsymbol{w}_{i_l}^\beta[\mu]})} {(\mathrm{g}^{\sum_{i_{l=1}}^{ i_L} s_{i_l}})^{u_{\ell_\beta}}} = \mathrm{g}^{\sum_{i_{l=1}}^{ i_L} \boldsymbol{w}_{i_l}^\beta[\mu]}}
\end{align}
The $\mathcal{AS}$ then utilizes a discrete logarithm method, such as {\em Pollard's rho algorithm}, to compute the aggregated model parameter $\sum_{i_{l=1}}^{i_L}\boldsymbol{w}_{i_l}^\beta[\mu]$. The average value is then calculated by the $\mathcal{AS}$, and the global model parameter $\boldsymbol{w}^\beta[\mu]$ is updated accordingly. Finally, the $\mathcal{AS}$ sends the updated global model back, in plaintext,  to the satellites for subsequent training iterations (cf. \sref{sec:overview}); these steps continue until convergence.

\section{Performance Evaluation}\label{section 3}

\subsection{Experiment setup}

\textbf{LEO Constellation \& Communication Links.} We examine an LEO constellation that comprises 20 satellites divided into 4 orbits, all orbiting at an altitude of 1200 km with an inclination angle of 70$^\circ$. We consider a GS located in Rolla, Missouri, USA as the $\mathcal {AS}$ (it can be anywhere on Earth) with a minimum elevation angle of 10$^\circ$. We employ a Systems Tool Kit simulator developed by AGI company to determine satellite-GS connectivity. The communication parameters introduced in Section~\ref{Com_link} are configured as: $P$=40 dBm, $G_i$=$G_s$=6.98 dBi, $T$=354.81K, $B$=50 MHz, and $f$=2.5 GHz.  Satellites' connectivity with the $\mathcal {AS}$ (GS) was established over a two-day period to obtain the convergence results.

\textbf{Satellites Training.} We train our local models on LEO satellites using the DeepGlobe dataset \cite{demir2018deepglobe}, which includes 1146 high-resolution colored satellite images (2448×2448 pixels), divided into training/validation/test sets (803/171/172 images). The dataset covers a total area of 1716.9 km$^2$ and has a pixel resolution of 50 cm. Each image has a corresponding mask image {(which is the ``label'')} with annotations for land cover, including 7 classes represented in different colors: \textcolor{cyan}{Urban}, \mycontourrr  {Agriculture}, \textcolor{magenta} {Rangeland}, \textcolor{green} {Forest}, \textcolor{blue} {Water}, \mycontour{Barren}, and \mycontourr {Unknown}. We augment and split the dataset evenly among 20 satellites. Each satellite trains a DeepLabV3+ model using its assigned data with a mini-batch size of 4, and a learning rate $\zeta$=0.00008.

\textbf{Baselines.} We compare FedSecure with most recent FL-LEO approaches including FedISL \cite{razmi}, FedHAP \cite{elfedhap}, and FedSpace \cite{so2022fedspace} in terms of convergence speed and accuracy. Note that FedSecure is the only approach that incorporates encryption.

\vspace{-0.1cm}
\subsection{Security and Privacy Analysis}

FedSecure is a secure FL aggregation approach that eliminates the need for a trusted KDC. Thus, it is much more resilient than FE-based methods~\cite{abdalla2018multi} (including IPFE) which require a KDC and a secure channel for generating and distributing secret keys to other nodes. Despite that, FedSecure achieves the same level of security as FE/IPFE-based methods. In the following, we explain how FedSecure protects privacy against the threat model discussed in Section~\ref{Threat}:

\begin{itemize}
    \item Although eavesdroppers may be able to intercept the exchanged ciphertexts, they will not be able to gain access to any information relevant to either the local model parameters of LEO satellites or the training data. This is because the model parameters are encrypted by the satellites using their own secret keys, making it impossible for anyone (including $\mathcal{AS}$) to decipher the parameters.

    \item Using FedSecure, the $\mathcal{AS}$ can only obtain the updated global model parameters using the aggregation key $AK_\mathcal{AS}$ given the encrypted model parameters received from the LEO satellites, but is not able to compute the individual model parameters. Moreover, despite possessing the subset aggregation key $S_{ {i_l}}$ of each satellite, it is unable to obtain their secret keys $s_{i_l}$ due to being masked by $\mathrm{g}^{\mathrm{x}_{{i_l}}\mathrm{y}_{{i_l}}}$.

    \item FedSecure is resilient against collusion attacks that may be launched among satellites that are owned by different operators 
    since they do not have access to the secret keys of non-colluding satellites, thus preventing them from decrypting their ciphertexts.
\end{itemize}
Thus, FedSecure ensures that sensitive satellite models remain secure even in the face of potential internal and external attacks.
 \vspace{-5mm}
\subsection{Computation \& Communication Overhead Analysis}

\subsubsection{Computation overhead}
We evaluate the computation overhead by measuring the time it takes for the $\mathcal {AS}$ and satellites to compute weights. We use the Python Charm library to implement our IPFE-based encryption scheme and run it on a 64-bit Ubuntu-based standard desktop computer with 4GB RAM and an Intel Core I3 CPU operating at 1.8GHz. 
According to Eq.~\ref{Enc_eq}, each satellite in each FL round encrypts $e$ elements including weights and biases. Our measurement shows that this computation takes less than 9 ms only, which is dominantly driven by a single exponentiation operation.

\subsubsection{Communication overhead}
The communication overhead of FedSecure is evaluated by examining the size and number of messages transmitted between the $\mathcal {AS}$ and the satellites. We employ a 160-bit security-level elliptic curve for the cryptographic operations in our scheme. Each satellite transmits encrypted matrices that represent its updated local model parameters (weights and biases) with a total number of elements $e$. In accordance with Eq.~\ref{Enc_eq} and given the DeepLabV3+ model structure, each satellite sends a total of $992$ MB of data in each FL round. However, it has been demonstrated by~\cite{king2004point} that elliptic curve points can be condensed, hence resulting in a smaller number of bits. 
This consequently leads to a communication overhead of $497$ MB using our scheme.

\subsection{Efficiency and Convergence Analysis}
We use the following performance metrics: 

\begin{itemize}[leftmargin=*]
\item {\bf Intersection over Union:} IoU provides a pixel-wise score for each class of objects in an image, ranging from 0 to 1. A score of 1 denotes a perfect match between the predicted and ground truth masks, while 0 indicates no overlap. The IoU score can be calculated as
\begin{equation} 
    IoU_b=\frac{\sum_{a=1}^{r} TP_{ab}}{\sum_{a=1}^{r} TP_{ab} + \sum_{a=1}^{r} FP_{ab} + \sum_{a=1}^{r} FN_{ab}}
\end{equation}
where $TP_{ab}$ and $FP_{ab}$ are the number of truly predicted and falsely predicted pixels as class $b$ in image $a$, respectively, $FN_{ab}$ is the number of falsely predicted pixels as other classes in image $a$ except for class $b$, and $r$ is the total number of images. Assuming $g$ land cover classes, the final score is the average of IoU across all classes, which can be expressed as
\vspace{-2mm}
   \begin{equation}
        mIoU=\frac{1}{g} \sum_{b=1}^{g} IoU_b
    \end{equation} \vspace{-1mm}
\item  {\bf Dice Coefficient:} Similar to IoU, this also measures the match between the predicted and the ground truth segmentation for each class, ranging from 0 (no overlap) to 1 (perfect overlap). But the Dice score places more emphasis on true positive predictions compared to false positives and false negatives, which can be calculated as
\end{itemize}

\begin{equation}
Dice_b = \frac{2 IoU_b}{1+IoU_b}
\end{equation}
$Dice$ can be also averaged among all classes to obtain $mDice$, just like $mIoU$ in the above.

Note that in all cases, the classification output is a semantic segmentation mask encoded in RGB format, where each pixel's color represents its class.


\begin{table}[!h]
\setlength{\tabcolsep}{0.45em}
\centering
\renewcommand{\arraystretch}{1.3}
\caption{Performance of FedSecure.} 
\label{table_comp}
 \begin{tabular}{|p{1.5cm}|p{1.1cm} | p{1.1cm}| p{1.1cm}| p{1.1cm} | p{1.1cm} |}
 \hline
   \centering Evaluation  &\multicolumn{5}{c|} {Communication round \& accumulative time (h)} \\
 \cline{2-6}
\centering Metric &\rmfamily 1 (1.51 h)& 2 (1.73 h)& 3 (2.18 h)&4 (2.76h)& 5 (2.97 h)\\
 \hline 
  {mIoU} &0.77896& 0.78016& 0.78075& 0.78148&0.78248 \\
    \cline{1-6}
 {mDice} &0.85141& 0.85192& 0.85259& 0.85308&0.85350  \\
 \hline 
\end{tabular}\vspace{-1mm}
\end{table}

In Table~\ref{table_comp}, we present FedSecure's evaluation results in terms of mIoU and mDice for the first five global epochs. In the first round, FedSecure achieved a high mIoU of 77.896\% and mDice of 85.141\%, indicating that it can accurately classify various target classes even after only one round ($\approx$1.5 hours). This faster convergence speed can be attributed to our partial aggregation scheme which enables all satellites' models on the same orbit to be aggregated before being encrypted and sent to the $\mathcal {AS}$. Subsequently, the precision of the global model increases gradually during the following communication rounds until it converges. 

\begin{figure}[!h]
\centering
    { \includegraphics[width=0.45\textwidth,height=1.9cm]{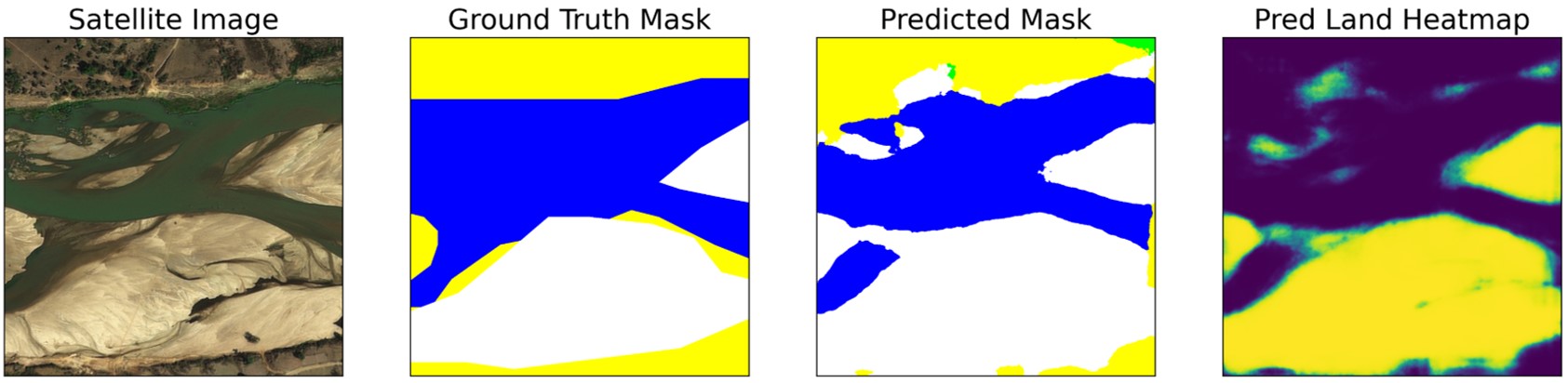}}
    { \includegraphics[width=0.45\textwidth,height=1.9cm]{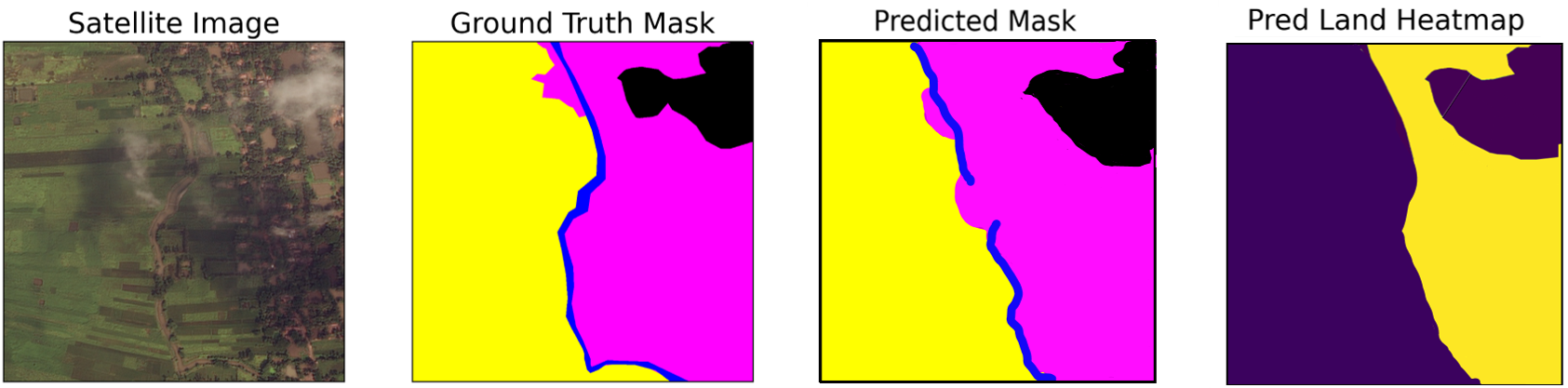}}
    { \includegraphics[width=0.45\textwidth,height=1.9cm]{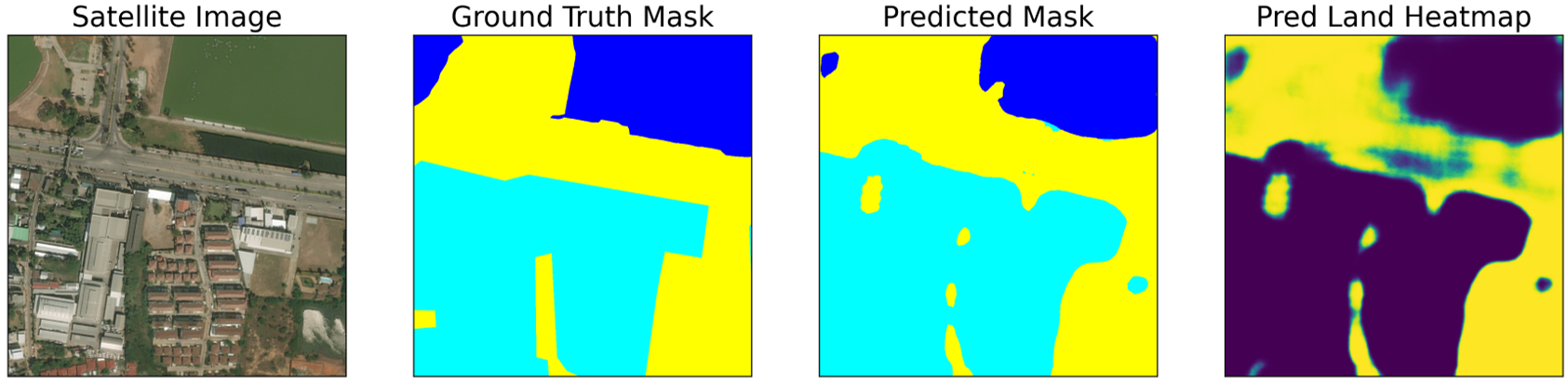}}
    { \includegraphics[width=0.45\textwidth,height=1.9cm]{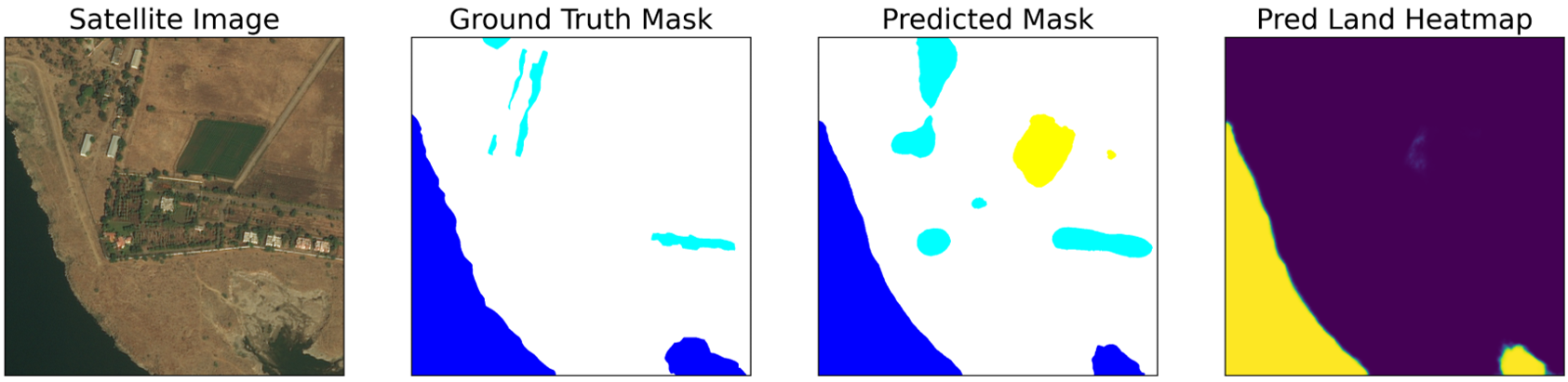}}
{\caption{Four examples showing the original satellite images, ground truth masks, predicted masks, and predicted land heatmap, by FedSecure after 3 hours of training. \label{DeepGlobe}}}
\vspace{-0.1cm} 
\end{figure}

We also evaluated our global model's effectiveness in classifying various land covers by testing it on {\em unseen satellite images} from the DeepGlobe dataset. As shown in Fig.~\ref{DeepGlobe}, the results indicate that after just three hours of satellites-$\mathcal {AS}$ communication,  FedSecure successfully predicted masks with a high rate of overlapping/matching with the ground truth masks. This result further demonstrates the effectiveness of FedSecure in achieving fast and high performance within a few hours. Moreover, we also note that the predicted masks can be improved further by allowing additional communication rounds.

\textbf{Comparison with baselines.} We compare FedSecure's convergence speed with the baselines using the MNIST dataset. Our results indicate that FedSecure converges within 3 hours with an accuracy of 88.76\% whereas FedISL \cite{razmi} archives 82.76\% accuracy after 4 hours of training. Notably, we achieve this level of accuracy when the $\mathcal{AS}$ (GS) is located in Rolla as a more practical configuration, not at the NP like FedISL. Moreover, when compared with FedHAP \cite{elfedhap} and FedSpace \cite{so2022fedspace} which require 15 and 96 hours to converge, respectively, our forwarding and aggregation scheme enables FedSecure to converge $\times$5 and $\times$32 faster, respectively. Most importantly, FedSecure ensures the security and privacy of satellite models, which is not addressed in those baselines.

\section{Conclusion} \label{Conclusion} 
This study proposes a novel FL-LEO framework, FedSecure, that offers secure and efficient model aggregation for distributed ML with satellites in the presence of attackers, eavesdroppers, and collusion. Unlike prior work, FedSecure eliminates the need for a key distribution center to generate private/public keys and the need for a secure channel to distribute the keys. Moreover, our on-orbit model forwarding and aggregation enables invisible satellites to participate in the FL process without waiting to become visible to the $\mathcal {AS}$, and does not require a sink satellite to collect models. Our simulation results demonstrate that FedSecure achieves the same level of security as existing methods while having low communication overhead (497 Mb), computation overhead ($<9$ms), and achieving fast convergence in only 3 hours on real satellite imagery dataset (DeepGlobe) with a classification accuracy of 85.35\%. 

\bibliographystyle{IEEEtran}
\renewcommand{\baselinestretch}{.9}
\small{
\bibliography{_references.bib}
}
\end{document}